\documentclass[prd,preprint,superscriptaddress]{revtex4}
\usepackage{revsymb}
\usepackage{amsmath}
\usepackage{graphicx}
\usepackage{bm}
\newcommand{\be}{\begin{equation}}
\newcommand{\ee}{\end{equation}}
\newcommand{\ben}{\begin{eqnarray}}
\newcommand{\een}{\end{eqnarray}}

\begin{document}
\title{Interacting models of soft coincidence}
\author{Sergio del Campo\footnote{Electronic Mail-address:
sdelcamp@ucv.cl}} \affiliation{Instituto de F\'{\i}sica,
Pontificia Universidad Cat\'{o}lica de Valpara\'{\i}so, Avenida
Brasil 2950, Casilla 4059, Valpara\'{\i}so, Chile}
\author{Ram\'{o}n Herrera\footnote{E-mail address: ramon.herrera.a@mail.ucv.cl}}
\affiliation{Instituto de F\'{\i}sica, Pontificia Universidad
Cat\'{o}lica de Valpara\'{\i}so, Avenida Brasil 2950, Casilla
4059, Valpara\'{\i}so, Chile}
\author{Germ\'an Olivares\footnote{E-mail address: german.olivares@uab.es}}
\affiliation{Departamento de F\'{\i}sica, Facultad de Ciencias,
Universidad Aut\'{o}noma de Barcelona, 08193 Bellaterra
(Barcelona), Spain}
\author{Diego Pav\'{o}n\footnote{E-mail address: diego.pavon@uab.es}}
\affiliation{Departamento de F\'{\i}sica, Facultad de Ciencias,
Universidad Aut\'{o}noma de Barcelona, 08193 Bellaterra
(Barcelona), Spain}

%%%%%%%%%%%%%%%%%%%%%%%%%%%%%%%%%%%%%%%%%%%%%%%%%%%%%%%%%%%%%%%%%%%%%%%%%%%%%
\begin{abstract}
The coincidence problem of late cosmic acceleration is a serious
riddle in connection with our understanding of the evolution of
the Universe. In this paper we show that an interaction between
the dark energy component (either phantom or quintessence) and
dark matter can alleviate it. In this scenario the baryon
component is independently conserved. This generalizes a previous
study [S. del Campo, R. Herrera, and D. Pav\'{o}n, Phys. Rev. D
\textbf{71}, 123529 (2005)] in which neither baryons nor phantom
energy were considered.
\end{abstract}

\maketitle

%%%%%%%%%%%%%%%%%%%%%%%%%%%%%%%%%%%%%%%%%%%%%%%%%%%%%%%%%%%%%%%%%%%%%%%%%%%%%%
\section{Introduction}
According to the conventional picture, the present accelerated
expansion of the Universe is driven by the negative pressure of an
unknown and unclustered component (dubbed ``dark energy") that
currently contributes about $70$\% of the total density. The
remaining $30$ per cent is shared between cold dark matter
($\rho_{dm} \sim 25$\%) and cold baryons ($\rho_{b} \sim 5$\%)
\cite{accel,wmap,hzt,snls,wmap3y}. The two latter, being
pressureless, redshift with expansion faster than the dark energy
component. Then, the cosmic coincidence problem arises, ``Why are
the densities of matter and dark energy  of precisely the same
order today?" \cite{coincidence}. Clearly, this is one outstanding
riddle in our understanding of the Universe. To solve it one is
forced to  adopt an evolving dark energy field (either
quintessence or phantom energy) or accept an incredibly  tiny
cosmological constant and admit that the ``coincidence" is just a
coincidence that hopefully might be somewhat alleviated with the
help of the anthropic idea \cite{reviews}. Here we take the view
that before resorting to the second option we should further
explore the first one. Yet, an evolving dark energy cannot solve
the problem either unless a suitable interaction (coupling) with
matter is allowed \cite{interaction,recent}. Note that the
coupling alters the rate at which  both matter and dark energy
redshift with expansion; this is why it can potentially alleviate
the aforesaid problem.

Interestingly, rather than in connection with the coincidence
problem, which was not even formulated at the moment, this
interaction was first proposed as a mechanism to reduce the value
of the cosmological constant \cite{wetterich}. Notice that in the
absence of underlying symmetry that would suppress the coupling
matter--dark energy there is no a priori reason to dismiss it. In
the last years, various proposals at the fundamental level for the
coupling leading to a constant ratio matter/dark energy at late
times were advanced (see e.g., \cite{federico} and references
therein), and specific phenomenological models have been built and
contrasted with observational data (high redshift supernovae and
CMB anisotropies) and seen to pass the tests \cite{tests}.
Further, the Akaike \cite{akaike} and Bayesian informative
criteria \cite{bic} when applied to high redshift supernovae data
suggest  a transfer of energy from the phantom component to the
matter component; yet, the conventional $\Lambda$CDM model is
still preferred \cite{szydlowsky}.

While a constant ratio matter/dark--energy at late times
(including the present one) would clearly alleviate the
coincidence problem it should be noted that a much less strong
condition would serve. It would suffice that nowadays the
aforesaid ratio varies only slowly (i.e., less faster than the
scale factor) and be of order unity (``soft coincidence").
Recently, it was found within this approach that the quintessence
scenario was favored over the tachyon scenario for the latter
would imply an excessive amount of pressureless matter today
\cite{recent}. However, in order to circumvent the tight
constraints on long--range forces \cite{tight} the baryon
component was left aside altogether. This may be justified because
the analysis was restricted to times about the present one which,
as said above, is characterized, among other things, by a low
value to the baryon density. In this paper we generalize the
analysis by including baryons in the energy density budget as an
independently conserved component and extend the study to earlier
times -though not to the radiation dominated era. As dark energy
components quintessence and phantom are separately considered.

The outline of the paper is as follows. In section II we present
the model. In section III we constrain it with recent high
redshift supernovae data. Finally, in section IV we discuss and
summarize our findings. As usual, a zero subscript or superscript
attached to any quantity means that it should be evaluated at the
present epoch.

%%%%%%%%%%%%%%%%%%%%%%%%%%%%%%%%%%%%%%%%%%%%%%%%%%%%%%%%%%%%%%%%%%%%%%%%
\section{The  interacting model}
We consider  a spatially flat Friedmann-Lemaitre-Robertson-Walker
universe dominated by a three-component system, namely, cold
baryonic matter, cold non-baryonic dark matter and dark energy,
such that the two latter components do not conserve separately but
interact with each other in a manner to be specified below. The
energy density and pressure of the dark energy, assuming it is a
quintessence field, are given by
\\
\begin{equation}
\rho_\phi = \textstyle{1\over{2}}\dot{\phi}^2 + V(\phi)\, , \qquad
\mbox{and} \qquad P_{\phi}  = \textstyle{1\over{2}}\ \dot{\phi}^2
- V(\phi), \label{rhop1}
\end{equation}
\\
respectively. If the dark energy is a phantom field we have
instead,
\\
\begin{equation}
\rho_\phi = -\textstyle{1\over{2}}\dot{\phi}^2 + V(\phi)\, ,
\qquad \mbox{and} \qquad P_{\phi}  = -\textstyle{1\over{2}}\
\dot{\phi}^2 - V(\phi). \label{phantom1}
\end{equation}
\\
The upper-dot stands for derivative with respect to the cosmic
time and $V(\phi)$ denotes both the quintessence field potential
and phantom potential. As is usually done, we postulate that the
dark energy component (either quintessence or phantom) obeys a
barotropic equation of state, i.e., $P_{\phi} = w_{\phi}\,%
\rho_{\phi}$ with $w_{\phi}$ a negative constant of order unity (a
distinguishing feature of dark energy fields is a high negative
pressure).

We assume that the dark matter and dark energy components are
coupled through a source (loss) term (say, $Q$) that enters the
energy balances
\\
\begin{equation}
\dot{\rho}_{dm} + 3 H \rho_{dm}  = Q ,
\label{cons1}
\end{equation}
\\
and
\\
\begin{equation}
\dot{\rho}_{\phi} + 3 H (\rho_{\phi} + P_{\phi}) = -Q \, .
\label{cons2}
\end{equation}
\\
In view of Eq.(\ref{rhop1}) last expression can alternatively
be written as $ \dot{\phi} \left[\ddot{\phi} + 3 H \dot{\phi} + V'%
\right] = - Q $. In  the case that the scalar field is of phantom
type the corresponding equation reads $\dot{\phi}%
\left[\ddot{\phi} + 3 H \dot{\phi} - V' \right] =  Q$, where the
prime denotes derivative with respect to $\phi$.

We consider that the baryon component is conserved whence its
energy density redshifts as
\\
\begin{equation}
\rho_{b} = \rho_b^{0}\left(\frac{a_{0}}{a}\right)^{3} .
\label{cons222}
\end{equation}

Defining $\rho_{m}=\rho_{b}+\rho_{dm}$ and using Eqs.(\ref{cons1})
and (\ref{cons222}) we obtain
\begin{equation}
 \dot{\rho}_{m} +3H \rho_{m}= Q .
\label{t2}
\end{equation}

Likewise, we assume that the interaction is related to the total
energy density of matter (baryonic plus dark) by $Q = 3 c^{2} H
\rho_{m}$, where $c^{2}$ is a small positive--definite constant.
As we shall see this choice ensures
that the ratio between the energy densities, $r(a)\equiv %
\rho_{m}/\rho_{\phi}$,  is a monotonous  decreasing function of
the scale factor, and such that around present time it varies very
slowly. By very slowly we mean that $\mid(\dot{r}/r)_{0}\mid \lesssim %
H_{0}$. This contrasts with previous studies in which $r$ was
demanded to asymptotically approach a fixed value at late times.

Clearly,
\\
\begin{equation}
\dot{r} \equiv
\frac{d}{dt}\left[\frac{\rho_{m}}{\rho_{\phi}}\right]=\frac{\rho_{m}}
{\rho_{\phi}}\left(\frac{\dot{\rho}_{m}}{\rho_{m}}-
\frac{\dot{\rho}_{\phi}}{\rho_{\phi}}\right).
\label{din}
\end{equation}
\\

\begin{figure}[th]
\includegraphics[width=3.5in,angle=0,clip=true]{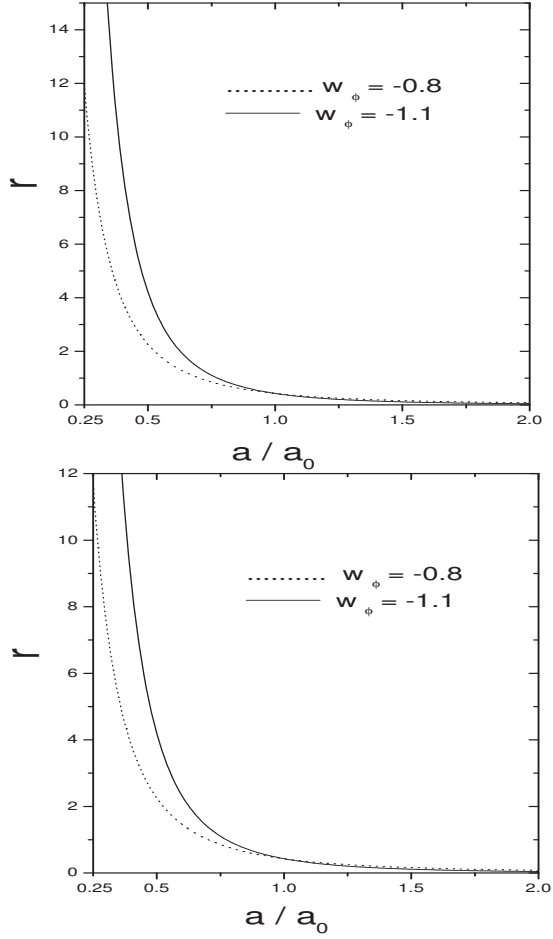}
\caption{Evolution of the ratio $\rho_{m}/\rho_{\phi}$ as given by
Eq.(\ref{r}) for $c^2 = 10^{-5}$ (upper panel) and $c^2 = 10^{-3}$
(lower panel). In both panels $r_{0} = 3/7$ with $w_{\phi}= -0.8$
(quintessence) and $w_{\phi}= -1.1$ (phantom). Not shown is the
behavior of $r$ for $a \rightarrow 0$ as the model does not apply
to such early times.} \label{fig:ratio1}
\end{figure}

In virtue of the above expressions last equation boils down to
\\
\begin{equation}
\dot{r} = 3 H r\left[ c^{2}(1+r) - |w_\phi| \right] \, ,
\label{cons6}
\end{equation}
\\
whose solution reads
\\
\begin{equation}
r=r_0 \,
\xi\left[c^2r_0-\left(\frac{a}{a_0}\right)^{3\xi}[c^{2}(1+r_{0})-|w_\phi|]
\right]^{-1}, \label{r}
\end{equation}
\\
with $\xi \equiv |w_{\phi}|-c^2 >0$. Figure \ref{fig:ratio1}
depicts the monotonous decrease of $r$ from values higher than
unity at early times to a nearly constant value at present time
(which we have fixed as $3/7$) irrespective of whether the dark
energy is a quintessence field or a phantom field. Notice that for
vanishing scale factor $r$ tends to the finite, constant value
$\xi/c^{2}$. However, our model should not be extrapolated to such
early stage.

One may wonder about the size of $c^{2}$. Obviously it should not
be large. In any case it must be lower than $|w_\phi|/(1+r_{0})$;
otherwise, by Eq. (\ref{cons6}), the Universe would have been
dominated by the dark energy from the beginning of the expansion
and galaxies would not have come into existence. On the other
hand, it should not be very small for it would have a negligible
impact and our model would be hardly distinguishable (depending on
the value of $w_{\phi}$) from the standard quintessence or phantom
models.

Because of $c^{2}$ ought to be a small quantity -though not very
small-, the right hand side of Eq.(\ref{r}) always stays above
zero, becoming negligible only for $a \gg a_{0}$. The closer
$c^{2}$ is to $|w_\phi|/(1+r_{0})$, the more alleviated the
coincidence problem gets. In particular, for $c^{2}> [\mid
w_{\phi}\mid -1/3](1+r_{0})^{-1}$ the current rate,
$\mid(\dot{r}/r)_{0}\mid$, results lower than $H_{0}$ (bear in
mind that the corresponding rate in case the dark energy were just
the cosmological constant is $3 H_{0}$), whereby the criterion of
``soft coincidence" is satisfied and the coincidence problem gets
significantly alleviated.

From Eq.(\ref{t2}) along with the expression for $Q$ we get
\\
\begin{equation}
\rho_{m}=\rho_{m}^{0}\; \left(\frac{a_0}{a}\right)^{3(1-c^2)}\, .
\label{rhom}
\end{equation}
\\
Thus, the densities of dark matter and dark energy are given by
\\
\begin{equation}
\rho_{dm}=\left(\frac{a_0}{a}\right)^{3}\left[\rho_{m}^0\left(\frac{a_0}{a}\right)^{-3c^2}-\rho_b^0
\right],
\label{ma}
\end{equation}
\\
and
\begin{equation}
\rho_\phi=\frac{\rho_{m}^0}{r_0\xi}\left(\frac{a_0}{a}\right)^{3(1-c^2)}
\left[c^2r_0-\left(\frac{a}{a_0}\right)^{3\xi}[c^{2}(1+r_{0})-|w_\phi|]
\right]\label{rho1}\, ,
\end{equation}
\\
respectively. Obviously, the constant $c^{2}$ could be determined
if either $\rho_{m}$, given by Eq. (\ref{rhom}) above, or the
ratio $\rho_{b}/\rho_{m} \propto a^{-3c^{2}}$, were accurately
known at different redshifts. The above expression for $\rho_{dm}$
may suggest that this quantity becomes negative at small scale
factor. That is so; however, if one takes into account that
$\rho^{0}_{b}/\rho^{0}_{m}$ is about $0.2$, for reasonable values
of $c^{2}$, this only occur well in the radiation era, i.e.,
beyond the range of applicability of our model.

As Fig. \ref{fig:ratio1} shows, $r \geq 10$ at sufficiently early
times (e.g., at redshifts larger than, say, $3$). This is
consistent with the analysis of Caldwell {\em et al.}
\cite{earlyq} who found that at the epoch of last scattering ($z
\simeq 1,100$) as well as at the onset of structure formation ($z
\sim 10^{3}$) $\Omega_{\phi}$ (the energy density of the dark
energy in units of the critical density) should not exceed $0.1$.

On the other hand, from Eqs. (\ref{cons1}) and (\ref{cons222})
alongside the condition $Q>0$ it follows that the baryon density
never dominates the matter density (i.e., $\rho_{b}/\rho_{dm}< 1$
always). This is in keeping with the widely accepted scenario of
cosmic structure formation, in which after the last scattering the
baryonic matter freely fells into the deep potential wells created
by the dark matter. If $\rho_{b}$ were larger than $\rho_{dm}$ at
early times, then the aforesaid scenario could be spoiled.

\begin{figure}[th]
\includegraphics[width=3.0in,angle=0,clip=true]{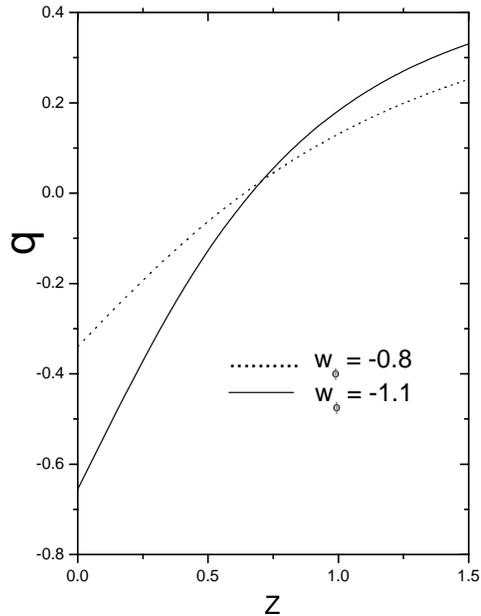}
\caption{The deceleration parameter as a function of the redshift,
$z= (a_0/a)-1$. Here $r_{0}=3/7$ and $c^2 = 10^{-3}$.}
\label{fig:deceleration}
\end{figure}

By combining Friedmann's equation
\\
\begin{equation}
3 H^2 = \kappa (\rho_b+\rho_{dm} + \rho_{\phi}) \quad \qquad(\kappa%
\equiv 8\pi\, G),
\label{Fried1}
\end{equation}
\\
with Eqs. (\ref{rhom})--(\ref{rho1}) we get the Hubble function
\\
\begin{equation}
H(a) = \frac{H_{0}}{\sqrt{(1+r_0)\xi}}
\,\left(\frac{a_0}{a}\right)^{\frac{3}{2}(1-c^2)} \left[|w_\phi|
\,r_0-\left(\frac{a}{a_0}\right)^{3\xi}[c^2(1+r_0)-|w_\phi|]
\right]^{1/2},
\end{equation}
\\
where $H_{0}= \sqrt{(1+r_0)\kappa \, \rho_{\phi}^{0}/3}$, which
will be needed below both to obtain the luminosity distance -a
previous step to draw the likelihood contours- and the
deceleration parameter, $q = -\ddot{a}/(aH^{2})$. The evolution of
the latter as a function of the redshift is depicted in Fig.
\ref{fig:deceleration}.

In this model the universe began accelerating only recently, at
redshifts about $0.65$. As we have checked numerically this
behavior is scarcely sensitive to the value of the $c^{2}$
parameter provided this lies in the range $10^{-5} \leq c^{2} \leq
10^{-1}$ (this is why we content ourselves with presenting just
two plots of $q$ vs. $z$). The larger $c^{2}$, the larger the
transition redshift.

Using Eq.(\ref{rhop1}) (respectively, Eq.(\ref{phantom1})) the
potential for the quintessence field  (respectively, phantom
field) in terms of the scale factor reads
\\
\begin{equation}
V(\phi(a))=(1+|w_\phi|)\;\frac{\rho_{m}^0}{
2r_0\xi}\left(\frac{a_0}{a}\right)^{3(1-c^2)}\left[c^2r_0-
\left(\frac{a}{a_0}\right)^{3\xi} [c^2(1+r_0)-|w_{\phi}|]\right].
\label{Vphia}
\end{equation}

To obtain the potential as a function of $\phi$ we must first
express the latter in terms of $a$. To this end, we write
\\
\begin{equation}
\dot{\phi}^2 = \pm
\left(\frac{d\phi}{da}\,H\,a\right)^2,\label{rdoop1}
\end{equation}
\\
where the plus(minus) sign  corresponds to quintessence field
(phantom field). And in virtue of Eqs. (\ref{rho1}),
(\ref{Fried1}) and (\ref{rdoop1}) it follows that
\\
\begin{equation}
\phi-\phi_0=\sqrt{\frac{\pm
3(1-|w_\phi|)}{\kappa}}\;\frac{1}{3\xi}\;\left[\left(\ln\Im(a)+\frac{c}{\sqrt{|w_\phi|}}\ln\Re(a)
\right)-C\right],
\end{equation}
\\
where
$$
\Im(a)=-c^2r_0+2c^2(a/a_0)^{3\xi}(1+r_0)-(r_0+2(a/a_0)^{3\xi})|w_\phi|
$$
$$
+2\sqrt{|w_\phi|(a/a_0)^{3\xi}
-c^2((a/a_0)^{3\xi}+r_0[(a/a_0)^{3\xi}-1]) }\,
\sqrt{|w_\phi|(r_0+(a/a_0)^{3\xi})-c^2(1+r_0)(a/a_0)^{3\xi}}\, ,
$$
\\
$$
\Re(a)=\frac{-3\xi(a/a_0)^{-3\xi}}{c^3
r_0\sqrt{|w_\phi|}}\;\;\left[\;c^{4}(1+r_0)(a/a_0)^{3\xi}+c^2r_0|w_\phi|[(a/a_0)^{3\xi}-2]
-w_\phi^2(a/a_0)^{3\xi}\right.
$$
$$
-2 c \sqrt{|w_\phi|} \,
\sqrt{|w_\phi|(a/a_0)^{3\xi}-c^2[(a/a_0)^{3\xi}+r_0((a/a_0)^{3\xi}-1)]}
$$
$$
\times\;\left.\sqrt{(r_0+(a/a_0)^{3\xi})|w_\phi|-c^2(1+r_0)(a/a_0)^{3\xi}}\right]\,
,
$$
\\
and $C$ denotes the integration constant
\[
C= \ln\Im(a_0)+\sqrt{\frac{c^{2}}{|w_\phi|}}\, \ln\Re(a_0)\, .
\]

\begin{figure}[th]
\includegraphics[width=3.0in,angle=0,clip=true]{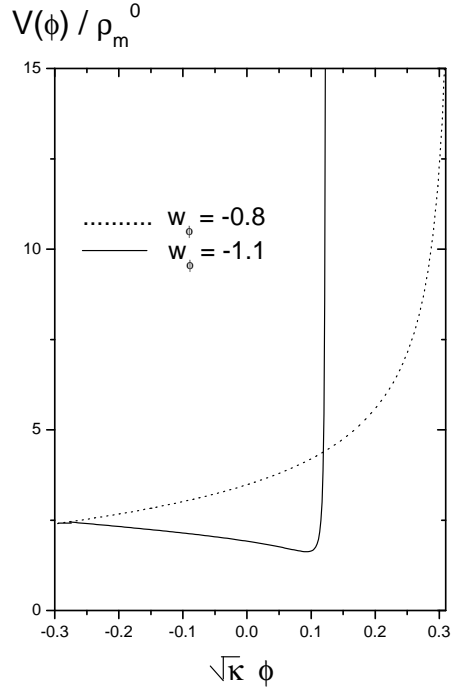}
\caption{ The normalized potential, $V(\phi)/\rho^{0}_{m}$,  vs
$\sqrt{\kappa}\phi$. Here $r_{0}=3/7$, $c^2= 10^{-3}$ and
$\phi_0=(C/3\xi)\;\sqrt{\pm 3(1-|w_\phi|)/\kappa}$, the plus
(minus) sign  corresponds to the quintessence field (phantom
field). } \label{fig:pot}
\end{figure}

The dependence of the potential on the field $\phi$ is shown in
Fig. \ref{fig:pot}.

Clearly, this potential must be understood as an effective one. It
should be noted that these effective potentials (quintessence and
phantom potentials) are similar to those used in inflationary
models, where the universe undergoes an accelerated period at very
early times. Thus, our potentials describe an accelerated phase at
present time, and simultaneously alleviate the coincidence
problem.

\section{Comparing with supernovae data}
In this section we use two independent supernovae type Ia (SNIa)
data sets, namely the ``gold" sample compiled by the High-Z
Supernovae Search Team (HZT) \cite{hzt}, and the  Supernova Legacy
Survey (SNLS) sample \cite{snls}, to constrain the parameters of
the model. The ``gold" sample, collected from different sources,
comprises $157$ SNIa, of redshifts up to $1.5$ ($14$ of which,
discovered by the Hubble Space Telescope, lie in the interval $1<
z <1.5$), with reduced calibration errors coming from systematics.
The SNLS sample is smaller, $71$ SNIa, with redshifts below unity.
However, the technique employed guarantees that no source is lost
and the data are of a higher quality.

\begin{figure}[th]
\includegraphics[width=4.5in,angle=0,clip=true]{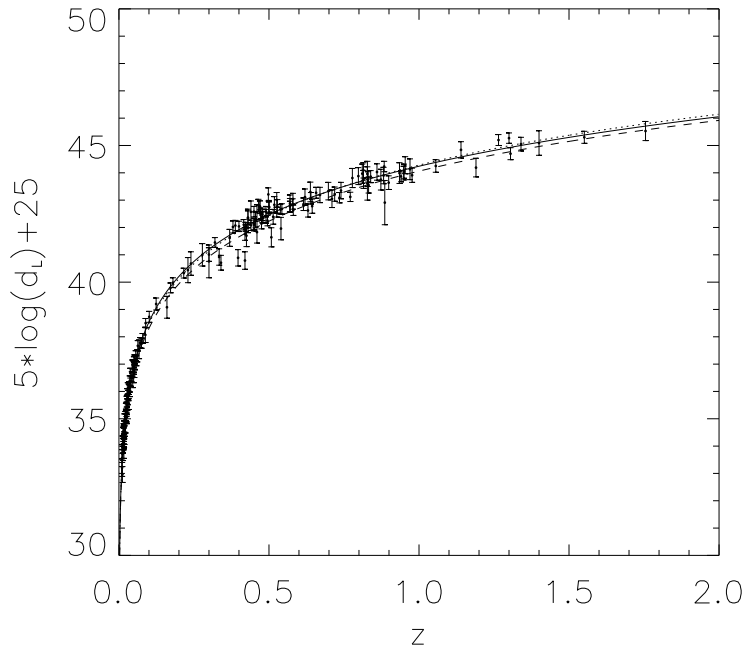}
\caption{Distance modulus  vs redshift for the best fit
quintessence model, $\Omega_{\phi} = 0.71$, $w_{\phi} = -0.99$,
$c^{2} = 10^{-5}$, $\chi^{2} = 177$ (dashed line), and the phantom
model, $\Omega_{\phi} = 0.40$, $w_{\phi} = -4.63$, $c^{2} = 0.20$,
$\chi^{2} = 173$ (solid line). The flat $\Lambda$CDM model is also
included, $\Omega_{\Lambda} = 0.70$, $w_{\Lambda} = -1$, $\chi^{2}
= 178$ (dotted line). The data points correspond to the ``gold"
sample of SNIa \cite{hzt}.} \label{fig:dL-HZT}
\end{figure}

\begin{figure}[th]
\includegraphics[width=4.5in,angle=0,clip=true]{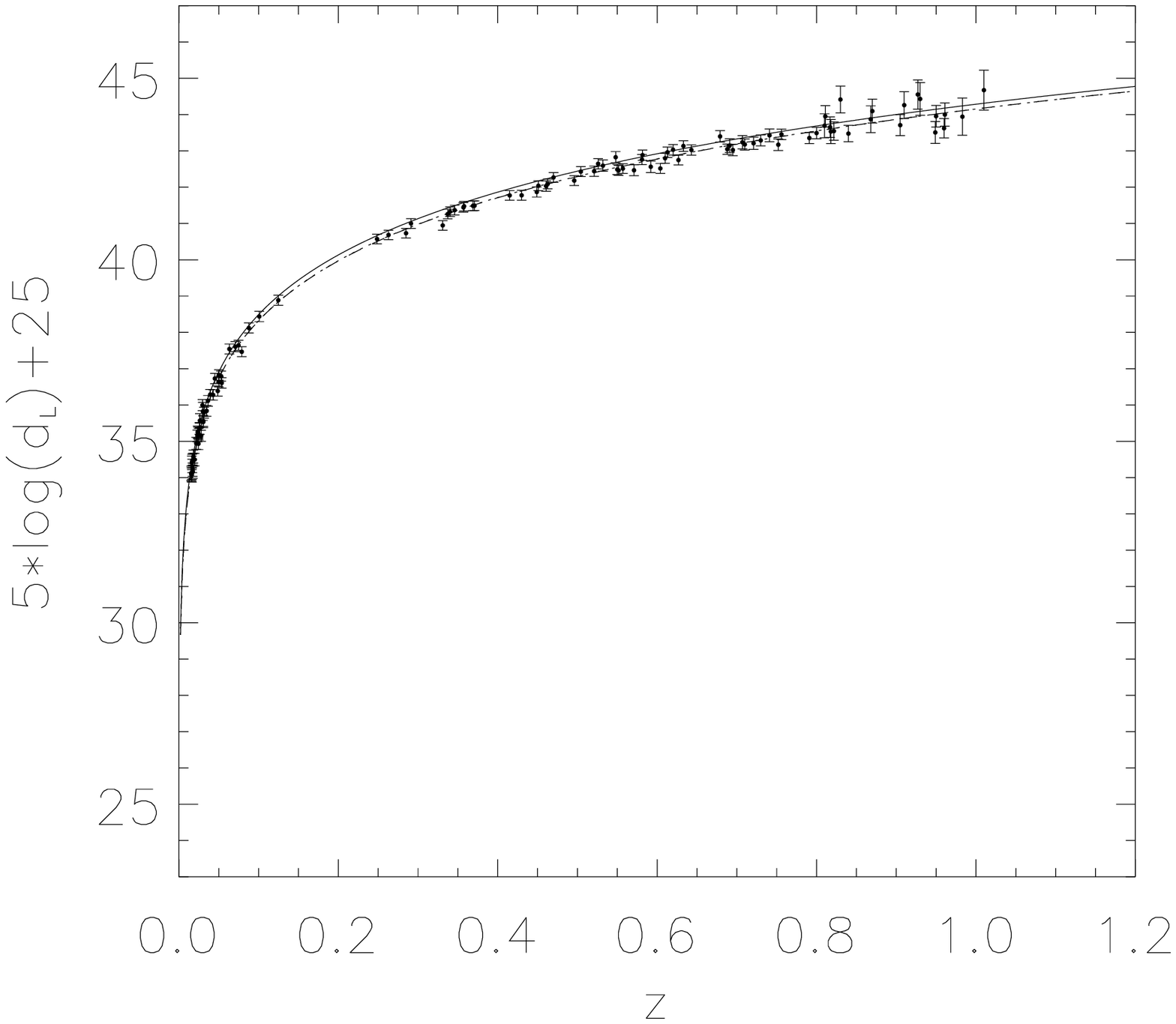}
\caption{Distance modulus vs redshift for the best fit
quintessence model, $\Omega_{\phi} = 0.74$, $w_{\phi} = -0.99$,
$c^{2} = 10^{-5}$, $\chi^{2} = 111.09$ (dashed line), and the
phantom model, $\Omega_{\phi} = 0.70$, $w_{\phi} = -1.10$, $c^{2}
= 0.035$, $\chi^{2} = 111.02$ (solid line). The flat $\Lambda$CDM
model is also included, $\Omega_{\Lambda} = 0.72$, $w_{\Lambda} =
-1$, $\chi^{2} = 111.03$ (dotted line). The data points correspond
to the SNLS sample \cite{snls}.} \label{fig:dL-SNLS}
\end{figure}

Figures \ref{fig:dL-HZT} and \ref{fig:dL-SNLS} depict the
best--fit of the phantom model (solid line) and the quintessence
model (dashed line) to the ``gold" sample and the SNLS sample,
respectively. For the sake of comparison the flat $\Lambda$CDM
model is also shown. In plotting
the graphs the expression for the distance modulus, $\mu = 5 \log%
d_{L} + 25$, was employed. Here $d_{L} = (1+z)%
\int_{0}^{z}{H^{-1}(z') dz'}$, is the luminosity distance in
megaparsecs.

\begin{figure}[th]
\includegraphics[width=6.5in,angle=0,clip=true]{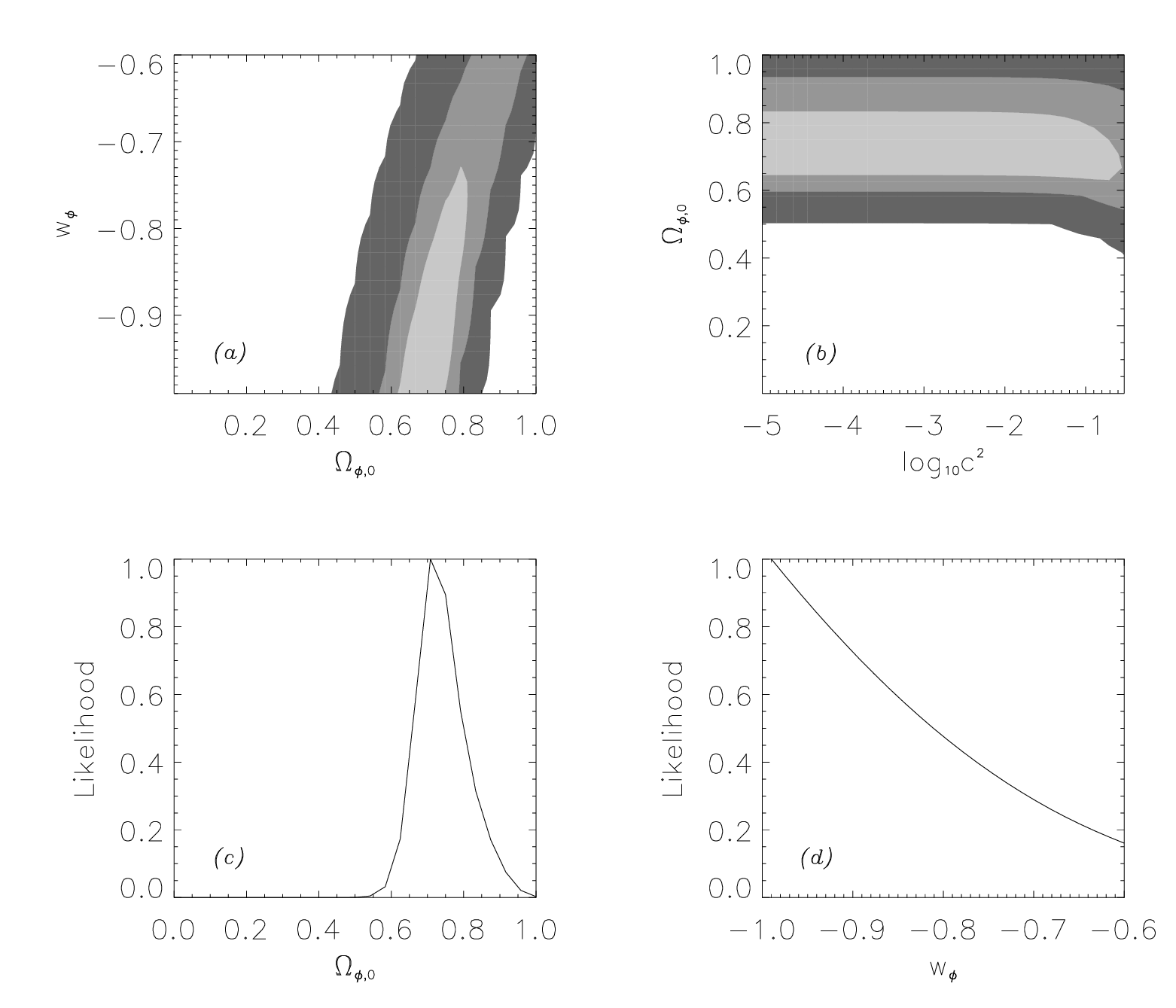}
\caption{Likelihood contours for the quintessence model displaying
the $68$\%, $95$\%, and $99.99$\% confidence intervals. The
likelihood are marginalized over the rest of the parameters. The
bottom panels show the the probability functions for the
quintessence energy density normalized to the critical density
(left panel), and the equation of state parameter of the
quintessence fluid (right panel). The data points correspond to
the ``gold" sample of SNIa \cite{hzt}.}
\label{fig:like-soft-quint-HZT}
\end{figure}

\begin{figure}[th]
\includegraphics[width=6.5in,angle=0,clip=true]{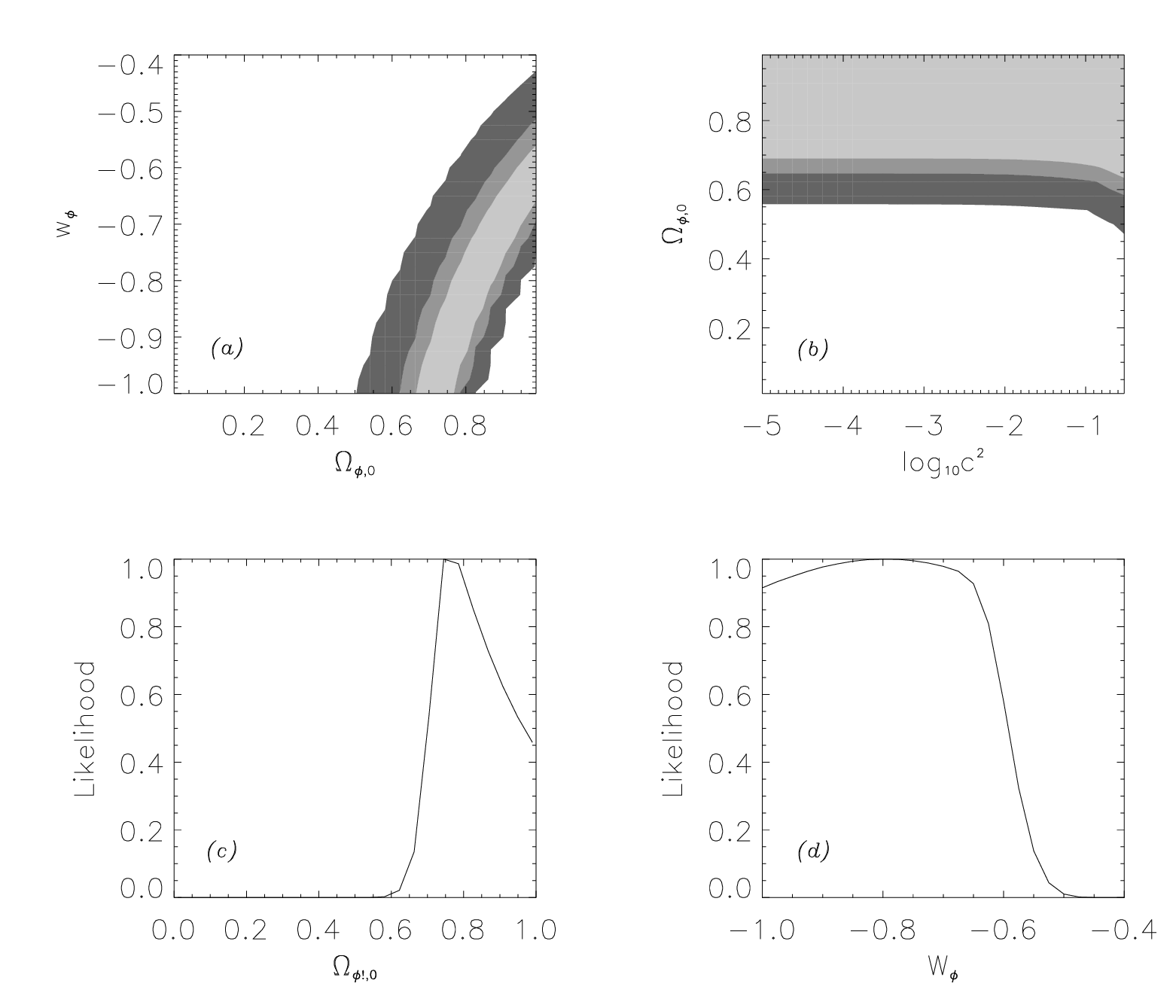}
\caption{Same as  Fig. \ref{fig:like-soft-quint-HZT} except that
the data points correspond to the SNLS data set of Ref.
\cite{snls}.} \label{fig:like-soft-quint-SNLS}
\end{figure}

Figures \ref{fig:like-soft-quint-HZT} and
\ref{fig:like-soft-quint-SNLS} portray the two--dimensional
likelihood contours for the case that the dark energy component is
a quintessence field, based on the ``gold" \cite{hzt} and the SNLS
sample \cite{snls}, respectively. Both set of contours were
calculated by running through a grid of models on a
four-dimensional parameter space. The prior $\Omega_{m} +
\Omega_{\phi} = 1$ was used and the present value of the Hubble
parameter was allow to vary in the interval $60 <H_{0}< 70\,
$km/s/Mpc. The rest of the priors are $10^{-5}<c^2<0.3$, and
$-1<w_\phi<-0.6$. The constraints obtained on the free parameters
from the HZT data are: $\Omega_{\phi} = 0.74^{+0.06}_{-0.07}$,
$w_{\phi}(1\sigma) < -0.9$, $c^{2}(1\sigma) < 0.13$. The latter
parameter shows large degeneracy and only in this case we find an
upper limit. In its turn, the constraints obtained on the free
parameters from the SNLS data are: $\Omega_{\phi} =
0.83^{+0.08}_{-0.09}$, $w_{\phi}(1\sigma) < -0.62$. Unfortunately,
the $c^2$ parameter is practically not constrained in this case.

\begin{figure}[th]
\includegraphics[width=6.5in,angle=0,clip=true]{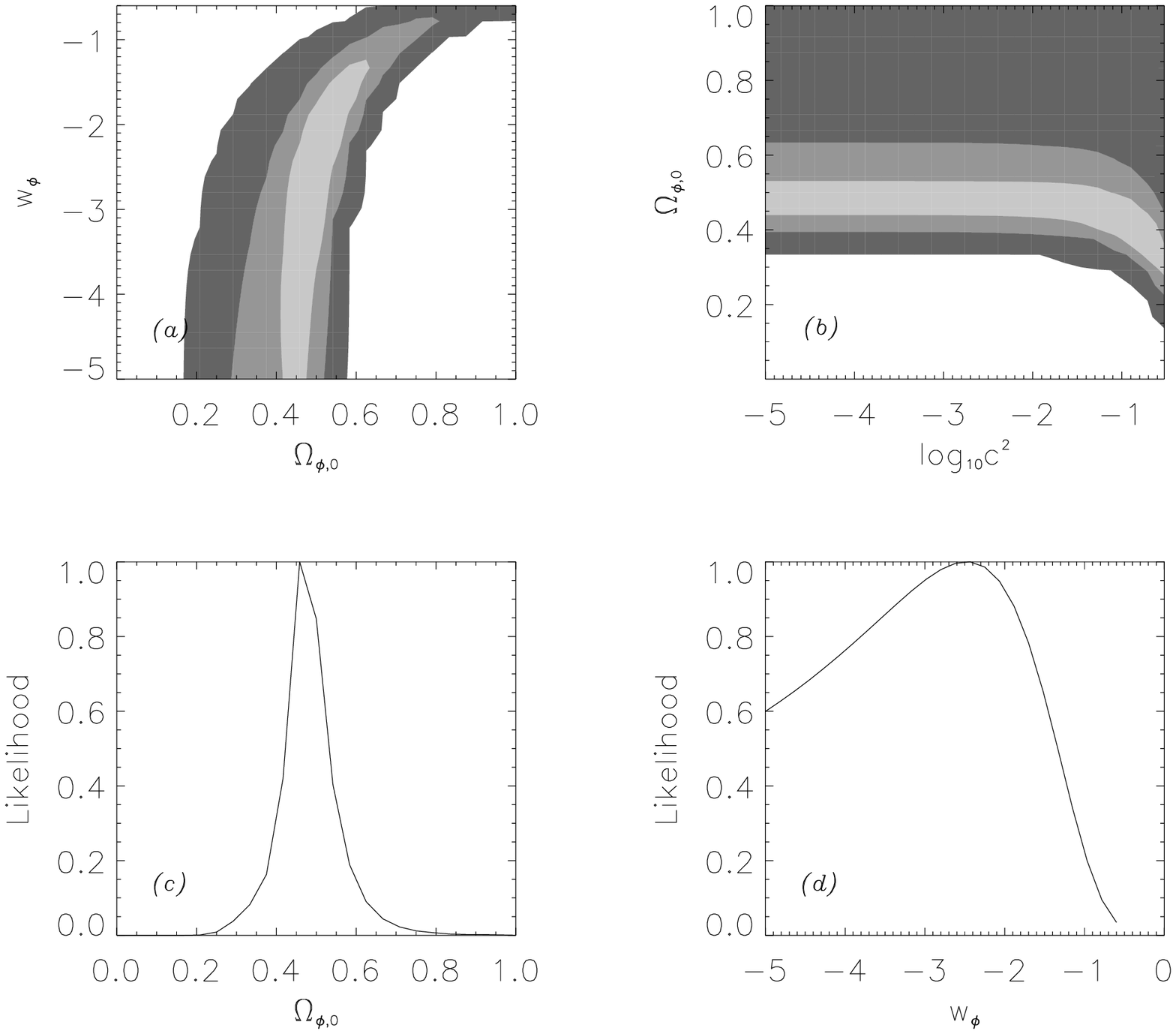}
\caption{Likelihood contours for the phantom model displaying the
$68$\%, $95$\%, and $99.99$\% confidence intervals. The likelihood
are marginalized over the rest of the parameters. The bottom
panels show the the probability functions for the phantom energy
density normalized to the critical density (left panel), and the
equation of state parameter of the phantom fluid (right panel).
The data points correspond to the ``gold" sample of SNIa
\cite{hzt}.} \label{fig:like-soft-phant-HZT}
\end{figure}

\begin{figure}[th]
\includegraphics[width=6.5in,angle=0,clip=true]{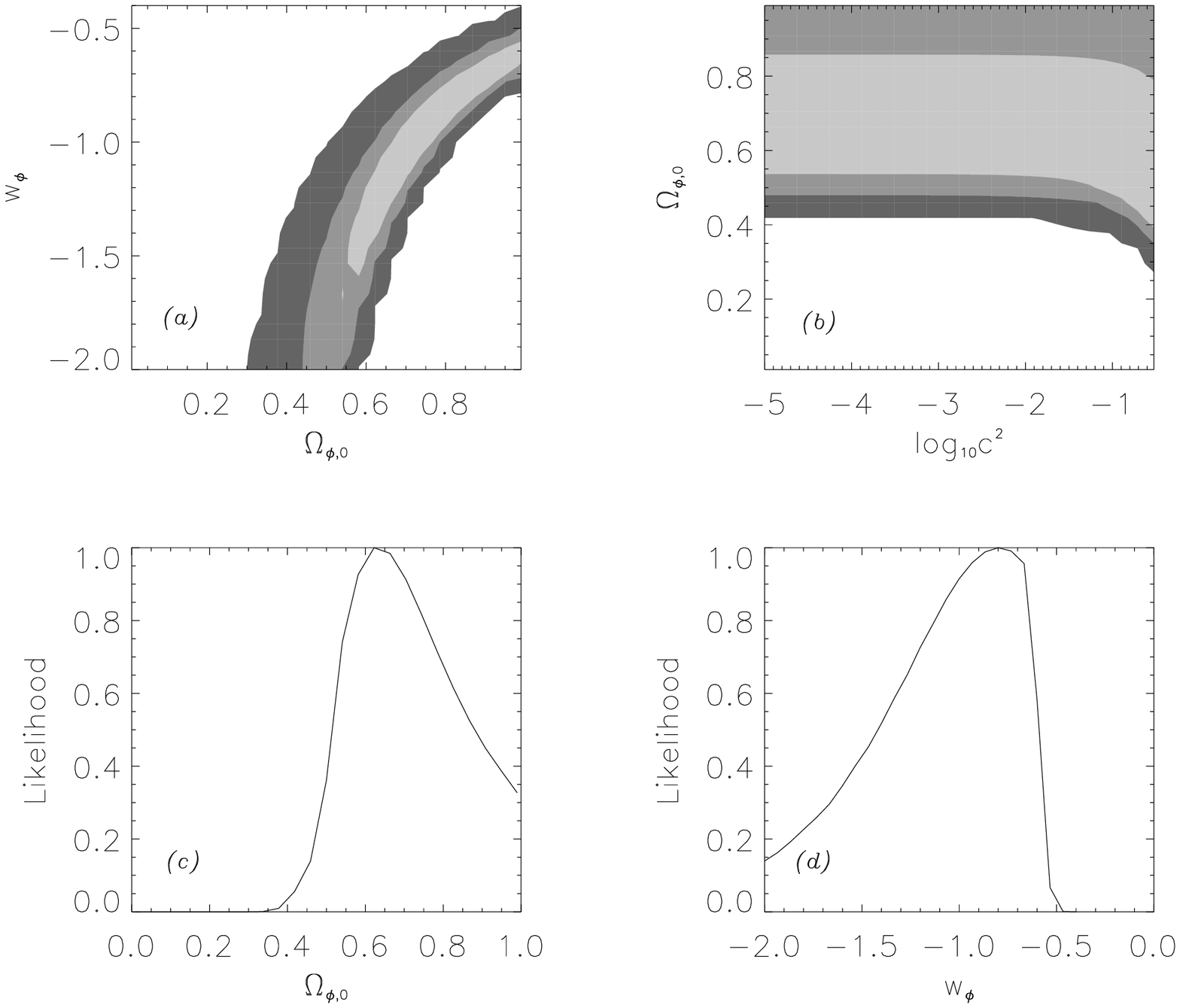}
\caption{Same as Fig. \ref{fig:like-soft-phant-HZT} except that
the data points correspond to the SNLS data set of Ref.
\cite{snls}.} \label{fig:like-soft-phant-SNLS}
\end{figure}

Similarly, Figs. \ref{fig:like-soft-phant-HZT} and
\ref{fig:like-soft-phant-SNLS} show the corresponding contours for
the phantom model using identical sets of data and the priors
$10^{-5}<c^2<0.3$, $-5 < w_{\phi} <-0.6$. The constraints obtained
on the free parameters from the first set (HZT) are:
$\Omega_{\phi} = 0.48 \pm 0.06$, $w_{\phi} = -3.0 \pm 1.1$. Again,
the interaction parameter $c^2$ shows a wide degeneracy.  In its
turn, the constraints obtained from the SNLS data set are: $\Omega_{\phi} =%
0.71^{+0.12}_{-0.13}$, $w_{\phi} = -1.1\pm 0.4$, with no real
constraint on $c^{2}$ given its large degeneracy.

As is usual in scenarios of late acceleration, this model prefers
a lower contribution of dark energy. The fact that phantom models
prefer a lower value of $\Omega_{\phi}$ is only natural given
their low value of $w_{\phi}$. It interesting to see that the SNLS
data favor a higher value for $\Omega_{\phi}$  (both for phantom
and quintessence models) than HZT's.

\section{Discussion}
We studied a model of late cosmic acceleration by assuming that
the dark matter and dark energy components are coupled to each
other so that there is a transfer of energy from the latter to the
former, while the baryon component is conserved. By suitably
choosing the interaction term, $Q$, the ratio between both dark
sources of gravity is seen to evolve, at present time, less faster
than the scale factor. This considerably alleviates the
coincidence problem albeit it does not solve it in full. Clearly,
to achieve the latter one should derive the present value of the
aforesaid ratio, or at least show that it has to be of order
unity. For the time being, $r_{0}$ ought to be understood as an
input parameter. This also holds for a handful of key
observational quantities such as the present value  of the cosmic
background radiation temperature, the cosmological constant
$H_{0}$, or the ratio between the number of baryons and photons.

The transition deceleration--acceleration occurs recently, at
redshifs lower than unity (see Fig. \ref{fig:deceleration}). This
contrasts with other interacting models in which the transition is
predicted to take place much earlier, at redshifhts as hig as $10$
\cite{transition}.

It is noteworthy that the expression for the potential, Eq.
(\ref{Vphia}), is identical irrespective of whether the dark
energy component is a phantom or a quintessence field.

Both when the dark energy is a quintessence or a phantom field the
model fits rather well the HZT data set, and in the latter case
clearly better than the concordance model $\Lambda$CDM does
($\chi^{2}_{phantom} = 173$, $\chi^{2}_{quintessence} = 177$,
$\chi^{2}_{\Lambda} = 178$). Yet, the Bayesian information
criterion (BIC) \cite{bic}, given by the formula BIC$= \chi^{2} +%
p\, \ln N$, where $p$ is the number of free parameters of the
model ($2$ for the $\Lambda$CDM model, $4$ for dark energy models)
and $N$ the number of data points, distinctly favors the
$\Lambda$CDM model for it yields a lower figure
(BIC$_{quintessence} \simeq 197$, BIC$_{phantom} \simeq 193$,
BIC$_{\Lambda} \simeq 188$).

The fits to the SNLS data are rather similar ($\chi^{2}_{phantom}
= 111.02$, $\chi^{2}_{quintessence} = 111.09$, $\chi^{2}_{\Lambda}
= 111.03$). However, the Bayesian information criterion again
sides with the $\Lambda$CDM model (BIC$_{quintessence} \simeq $
BIC$_{phantom} \simeq 128$, BIC$_{\Lambda} \simeq 119$).

We have not used the  Akaike criterion because it favors models
with larger number of parameters \cite{favors}.

Thus, from the point of view of the Bayesian information criterion
the $\Lambda$CDM model is  preferred, and  the soft coincidence
model for quintessence seems clearly disfavored if not discarded.
As for the phantom model, the situation appears somewhat
undecided. While, regarding the BIC, it lags behind the
concordance model it shows a better fit to the HZT sample of
supernovae and alleviates the coincidence problem. We may
summarize by saying that the SNIa data are not yet conclusive
thereby a richer supernovae statistics, especially at redshifts
larger than unity, is needed to come to a verdict.

Unfortunately, the supernovae constraints on the $c^{2}$ parameter
is rather poor. This is coherent with the fact that the
interaction dark energy--dark matter affects the luminosity
distance only at third order in the redshift \cite{grg1-2}.
Nevertheless, we hope to be able to break the degeneracy by
submitting the model to further tests such as the CMB temperature
anisotropies and the distribution of matter at cosmological
scales. This will be the subject of a future work.

\acknowledgments{We are indebted to Winfried Zimdahl for helpful
comments on an earlier draft of this article. SdC was supported
from Comisi\'{o}n Nacional de Ciencias y Tecnolog\'{\i}a (Chile)
through FONDECYT Grants 1030469, 1010485 and 1040624 and by the
PUCV under Grant 123.764/2004. R. H. was supported by the
``Programa Bicentenario de Ciencia y Tecnolog\'{\i}a" through the
Grant ``Inserci\'on de Investigadores Postdoctorales en la
Academia" \mbox {N$^0$ PSD/06}. This work was partially supported
by the old Spanish Ministry of Science and Technology under Grant
BFM--2003--06033, and the ``Direcci\'{o} General de Recerca de
Catalunya" under Grant  2005 SGR 000 87}.

\end{document}